\documentclass[aps,prb,twocolumn,superscriptaddress]{revtex4}

\usepackage{graphicx}
\usepackage{epstopdf}
\usepackage{color}
\usepackage{lettrine}
\usepackage{type1cm}

\usepackage[font=sf,labelfont=bf,justification=raggedright]{caption}
\DeclareCaptionLabelSeparator{line}{ $|$ }
\begin{document}


\title{\textsf{Magnetic microscopy of topologically protected homochiral domain walls in an ultrathin perpendicularly magnetized Co film}}



\author{M.~J.~Benitez}
\thanks{These authors contributed equally to this work.}
\affiliation{School of Physics and Astronomy, University of Glasgow, G12 8QQ, United Kingdom }
\author{A.~Hrabec}
\thanks{These authors contributed equally to this work.}
\affiliation{School of Physics and Astronomy, University of Leeds, Leeds LS2 9JT, United Kingdom}
\author{A.~P.~Mihai}
\affiliation{School of Physics and Astronomy, University of Leeds, Leeds LS2 9JT, United Kingdom}
\author{T.~A.~Moore}
\affiliation{School of Physics and Astronomy, University of Leeds, Leeds LS2 9JT, United Kingdom}
\author{G.~Burnell}
\affiliation{School of Physics and Astronomy, University of Leeds, Leeds LS2 9JT, United Kingdom}
\author{D.~McGrouther}
\affiliation{School of Physics and Astronomy, University of Glasgow, G12 8QQ, United Kingdom }
\author{C.~H.~Marrows}
\affiliation{School of Physics and Astronomy, University of Leeds, Leeds LS2 9JT, United Kingdom}
\author{S.~McVitie}
\affiliation{School of Physics and Astronomy, University of Glasgow, G12 8QQ, United Kingdom }


\begin{abstract}
  Next-generation concepts for solid-state memory devices are based on current-driven domain wall propagation, where the wall velocity governs the device performance. It has been shown that the domain wall velocity and the direction of travel is controlled by the nature of the wall and its chirality. This chirality is attributed to effects emerging from the lack of inversion symmetry at the interface between a ferromagnet and a heavy metal, leading to an interfacial Dzyaloshinskii-Moriya interaction that can control the shape and chirality of the magnetic domain wall. Here we present direct imaging of domain walls in Pt/Co/AlO$_x$ films using Lorentz transmission electron microscopy, demonstrating the presence of homochiral, and thus topologically protected, N\'{e}el walls. Such domain walls are good candidates for dense data storage, bringing the bit size down close to the limit of the domain wall width.
\end{abstract}

\date{\today}


\maketitle

\lettrine[lines=3]{\textcolor[gray]{0.5}{\textsf{T}}}{}he broken inversion symmetry at interfaces between ferromagnets and heavy (high spin-orbit interaction) metals offers new ways to manipulate the magnetic state. The combination of a heavy metal and a thin ferromagnetic film gives rise to new phenomena which normally vanish in the bulk, but play an important role as soon as the thickness of the ferromagnet is reduced to an atomic size. The presence of a Rashba electric field\cite{Miron_Nature2010} and Dzyaloshinskii-Moryia interaction\cite{chen2013tailoring} (DMI) have been recently demonstrated in such systems. Therefore the classical picture of magnetism as being an interplay between exchange, dipolar, and anisotropy energies is perturbed by a new energy term with very significant and profound consequences. This DMI term is expressed as $\mathbf{D}_{i,j}\cdot\left(\mathbf{S}_i \times \mathbf{S}_j\right)$, where $\mathbf{D}_{i,j}$ is the DMI vector, and $\mathbf{S}_i$ and $\mathbf{S}_j$ are spin moments sitting on neighbouring atoms. When $\mathbf{D}$ is sufficiently strong, a non-uniform ferromagnetic state has a lower energy giving rise to exotic structures such as cycloids\cite{ferriani2008atomic}, helices\cite{rossler2006spontaneous}, or skyrmions \cite{heinze2011spontaneous} as the magnetic ground state: at an interface the DMI enforces spin textures of a cycloidal form\cite{fert1991magnetic,Crepieux1998,Freimuth2014}. However, for small DMI values, domain walls (DWs) are the precursors of these non-uniform states. Here the DMI strength is imprinted into the static DW texture via a virtual effective magnetic field that prefers a N\'{e}el wall of a given chirality rather than the magnetostatically cheaper Bloch wall. It has been shown that the choice of the heavy element dictates the sign and magnitude of the DMI\cite{ryu2014chiral, hrabec2014measuring, chen2013tailoring,torrejon2014interface}. The nature of the DW has remarkable consequences for the processes of DW dynamics\cite{de2013piezoelectric} and the sensitivity of the DW to the torques exerted on the localized magnetic moments\cite{khvalkovskiy2013matching}. The DW texture can be deduced indirectly by matching models with current-\cite{emori2013current,ryu2013chiral,conte2014interfacial} and field-induced\cite{Thiaville_DMI,choe,hrabec2014measuring} DW displacement data.

Determination of the DW structure is therefore a key part of any investigation in these materials. Historically, imaging of DWs in films with planar magnetization is well established with many examples demonstrating the capability of imaging DWs which have widths upwards of tens to hundreds of nm. Methods such as magnetic force microscopy\cite{yamaguchi2004real}, photoemission electron microscopy\cite{klaui2004head}, electron holography\cite{backes2007transverse}, or Lorentz transmission electron microscopy\cite{basith2011direct} (L-TEM) have been used to image such DWs. The spatial extent of DWs in materials with out-of-plane anisotropy is often only 10~nm or less, making them interesting objects for high density data storage devices. Resolving such small magnetic objects is a challenge but a number of methods, including those listed above, are suitable for such studies. Spin-polarized scanning tunnelling microscopy\cite{Kubetzka_2003} and spin-polarized low energy electron microscopy\cite{chen2013tailoring} manifested the N\'{e}el texture of the walls in bilayer systems grown and studied \textit{in situ}, while nitrogen-vacancy microscopy revealed and demonstrated the difference of the stray field distribution above the Bloch and N\'{e}el walls in trilayer systems such as Pt/Co/AlO$_x$ or Pt/Co/Pt\cite{tetienne2014nature}. Pt/Co/AlO$_x$ system is extremely interesting for solid state memory devices\cite{parkin2008magnetic} due to the combination of high DW velocities\cite{Moore_APL2008,moore2009erratum} (with various phenomena being proposed to explain this process\cite{Miron_Nature2010,haazen2013domain,Pizzini_DMI}) with the oxide layer which can be used as a tunnel barrier for the information read-out. Here we use L-TEM to directly image DWs in perpendicularly magnetized `device-ready' films of Pt/Co/AlO$_x$, allowing us to deduce the presence of narrow homochiral N\'{e}el walls. Furthermore, measurements taken using both L-TEM and polar Kerr imaging, which demonstrate the topological protection of these walls when they are forced together by a field, are consistent with the presence of a measurable DMI.

\section*{Results}

\begin{figure}[t]
  \includegraphics[width=8cm]{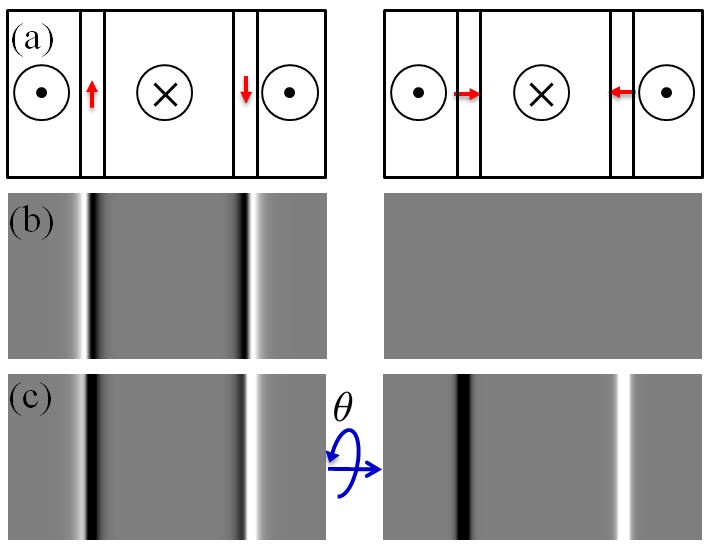}
  \caption{\textbf{Bloch (left) and N\'{e}el (right) domain wall structures in a perpendicularly magnetized thin film.} (\textbf{a}) Plan view sketch of Bloch (left) and N\'{e}el (right) walls in a thin film. The red arrows indicate the direction of the moments in the centre of the wall. The calculated Fresnel L-TEM images for these DWs are shown in (\textbf{b}) and (\textbf{c}) for cases with a sample tilt of  $\theta=0^{\circ}$ and $\theta=30^{\circ}$, respectively. \label{Fresnel_Contrast}}
\end{figure}

The two main types of DWs for films possessing out-of-plane anisotropy, Bloch and N\'{e}el, are sketched in figure~\ref{Fresnel_Contrast}(a) in plan view. The Bloch wall creates charges only at the top and bottom edges of the film, which are well separated for walls of any significant length, but is divergence-free otherwise. In contrast to this, the N\'{e}el wall gives rise to positive and negative magnetic charges along the whole length of the DW, separated by the DW width $\Delta$, which is of the order of $\Delta \approx (A/K)^{1/2}$, where $A$ is the exchange stiffness and $K$ is the anisotropy constant. Thus the Bloch wall is energetically favoured in perpendicularly magnetized thin films on magnetostatic grounds. To rotate the magnetic moments from the Bloch into the N\'{e}el configuration, an in-plane anisotropy expressed as $K_\mathrm{D} = N_x\mu_0M_\mathrm{s}^2/2$, where $N_x$ is the demagnetizing factor\cite{tarasenko} that depends on the DW width $\Delta$ and film thickness $t$, has to be overcome. This can be achieved either by applying an external magnetic field in the $x$-direction\cite{haazen2013domain}, or by an effective magnetic field arising from the DMI\cite{Thiaville_DMI}.

\begin{figure}[t]
  \includegraphics[width=8cm]{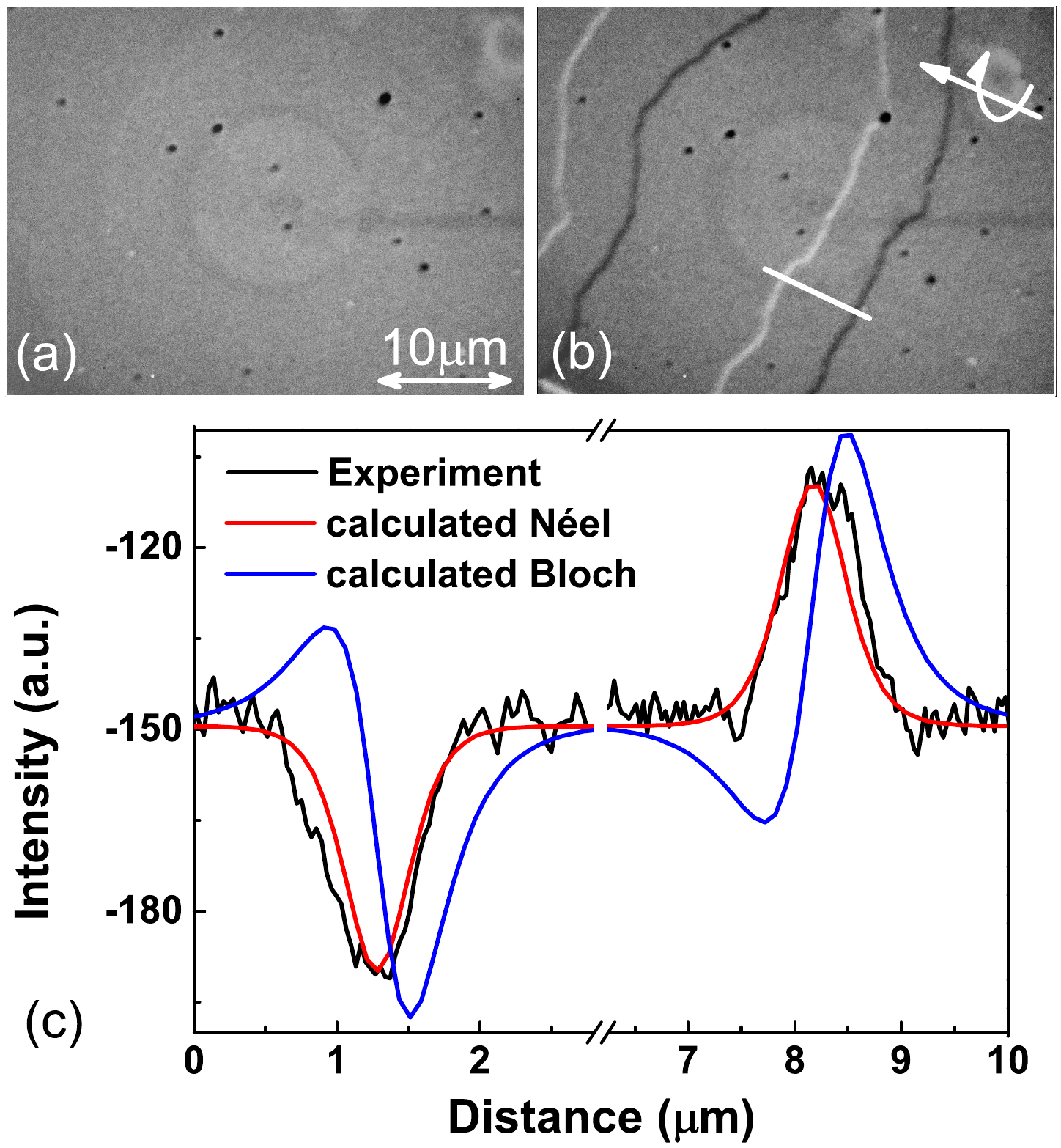}
  \caption{\textbf{Fresnel mode L-TEM images of the Pt/Co/AlO$_x$ sample.} (\textbf{a}) At normal incidence, $\theta=0^{\circ}$,the DWs generate no contrast and are invisible. (\textbf{b}) For a tilted sample, $\theta=30^{\circ}$, with the tilt direction indicated by the white arrows, the DWs appear. That the same area has been imaged is confirmed by the pattern of dark spots, which correspond to defects or dust particles. (\textbf{c}) Contrast linetraces: the experimental contrast was obtained at the position of the white line in (b). The calculated Bloch and N\'{e}el wall contrast was extracted from figure~ \ref{Fresnel_Contrast}(c). \label{Imaging}}
\end{figure}

To enable observation of the DWs, Fresnel-mode L-TEM imaging was carried out, which reveals DWs as lines of black/white contrast\cite{chapman1984investigation}. For materials with in-plane magnetization this is achieved by defocussing the imaging lens; the induction either side of the DW results in a Lorentz deflection and the electron beam diverges or converges at the wall giving dark or bright wall contrast. However, for perpendicularly magnetized films, no deflection arises from the out-of-plane component if the electron beam is at normal incidence. Thus any contrast will arise from the in-plane component, which only exists at the position of the DW. For the case of Bloch and N\'{e}el DWs, the calculated Fresnel images are shown in figure~\ref{Fresnel_Contrast}(b) for normal incidence. The Bloch wall on the left shows black/white contrast as each edge of the DW is similar a diverging or converging wall, but with no induction on one side of the wall, as has been observed experimentally \cite{masseboeuf2009lorentz}. In the case of N\'{e}el walls at normal incidence no contrast is visible, as shown in the right hand image of figure~\ref{Fresnel_Contrast}(b) calculated from the magnetization configuration. The reason for this is that the wall rotation is completely divergent in nature and gives no contrast in Lorentz microscopy. The basis of contrast in Lorentz microscopy is that the magnetization configuration must have a component of magnetization curl parallel to the electron beam direction \cite{mcvitie2006quantitative}; there is no such component present in the N\'{e}el wall. This distinguishes the two walls, although the N\'{e}el walls are invisible in this orientation.

The N\'{e}el wall can be made visible by tilting the film along an axis perpendicular to the wall in the plane of the film so that the electron beam no longer passes through the film at normal incidence. This then results in a component of magnetization/induction in the domains that is perpendicular to the beam and gives only black or only white contrast at the walls, as shown in figure~\ref{Fresnel_Contrast}(c) on the right hand side. Tilting of a film with a Bloch wall results in contrast observed in figure~\ref{Fresnel_Contrast}(c) on the left, which retains its black/white character although now in an asymmetric form. 
In this way the wall positions are identified and their forms distinguished. In summary, Bloch walls will be visible when the film is untilted and show asymmetric black/white contrast when tilted, whilst N\'{e}el walls will be invisible when untilted but show symmetric black or white contrast when tilted.

\begin{figure*}[t]
  \includegraphics[width=17cm]{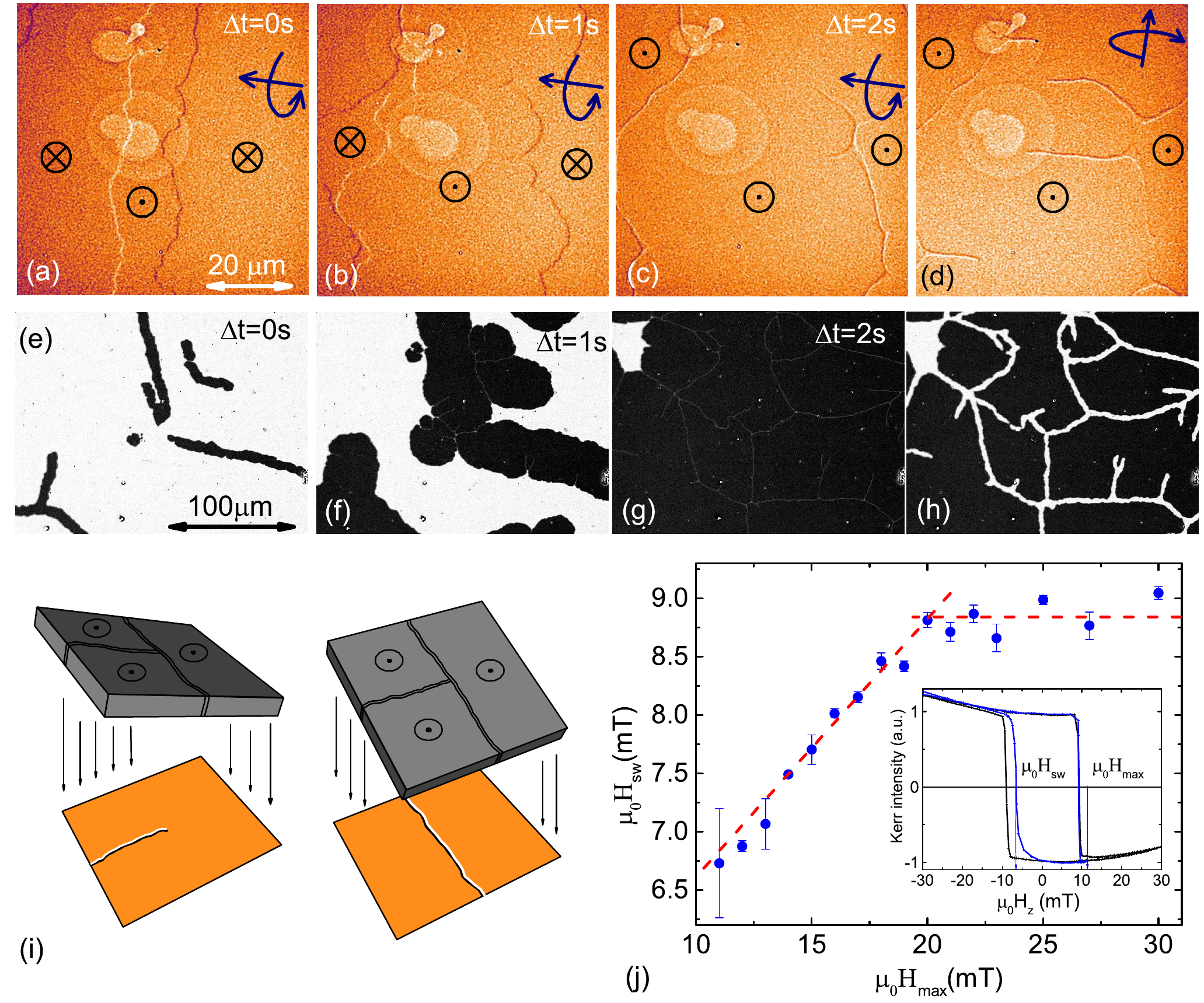}
  \caption{\textbf{Field-driven DW displacement experiments.} (\textbf{a})-(\textbf{c}) L-TEM images of the DWs displaced by external field of $B_{\mathrm{ext}}=2$~mT as a function of time. (\textbf{d}) The character of the N\'{e}el walls is confirmed by tilting the sample as indicated by the arrows. (\textbf{e})-(\textbf{g}) Polar Kerr microscopy images showing the magnetization reversal process, with topologically protected 360$^\circ$ DW structures indicated by the white lines between domains at 7~mT (panel (\textbf{g})). Therefore they behave as artificial nucleation centers when the polarity of the external magnetic field is reversed in (\textbf{h}). (\textbf{i}) The contrast from two perpendicular DWs can be revealed by tilting the sample with the respect to the DW orientation.  (\textbf{j}) Switching field $\mu_0H_{\mathrm{sw}}$ as a function of $\mu_0H_{\mathrm{max}}$. The dashed lines correspond to the two linear regimes. Inset shows hysteresis loop cycled between $\pm30$~mT (black) and $+30$~mT and $\mu_0H_{\mathrm{max}}$ (blue).\label{DW_squashing}}
\end{figure*}

The walls under investigation here in sputtered Pt/Co/AlO$_x$ trilayers were imaged in the Fresnel mode, and the results are shown in figure~\ref{Imaging}. First, the film was subjected to an out-of-plane applied field of $\sim8$~mT  using the objective lens, close to value of the coercivity, to induce DWs. After the applied field had been removed, Fresnel images were taken both untilted and tilted by 30 degrees, which are shown in figure~\ref{Imaging}(a) and (b), respectively. It is apparent that no DWs at all are visible when the film is untilted, furthermore when tilted the DW contrast appears either dark or bright without any dark/bright asymmetry. An intensity linetrace across bright and dark walls is shown in figure~\ref{Imaging}(c) confirming the symmetric nature of the contrast. (A small intensity variation across the image can be seen in both images which is non magnetic in origin). To illustrate the difference between Bloch and N\'{e}el walls, calculated intensity variations derived from the tilted images in figure~\ref{Fresnel_Contrast}(c) are shown in figure~\ref{Imaging}(c) for comparison with the data. The clear asymmetry expected from the Bloch walls is not visible in the experimental linetrace. We conclude therefore that the walls are of the N\'{e}el type. It should be noted that the component of the magnetization at the centre of the N\'{e}el wall gives no contrast in Fresnel images in this configuration. Therefore, whilst we can conclude that these are indeed N\'{e}el walls, it is not possible to say anything about the chirality from such images alone. We now turn our attention to the domain wall behaviour in the presence of an applied field to determine whether chiral effects are induced by the asymmetric interfaces of the Co layer.

\begin{figure}[t]
  \includegraphics[width=8cm]{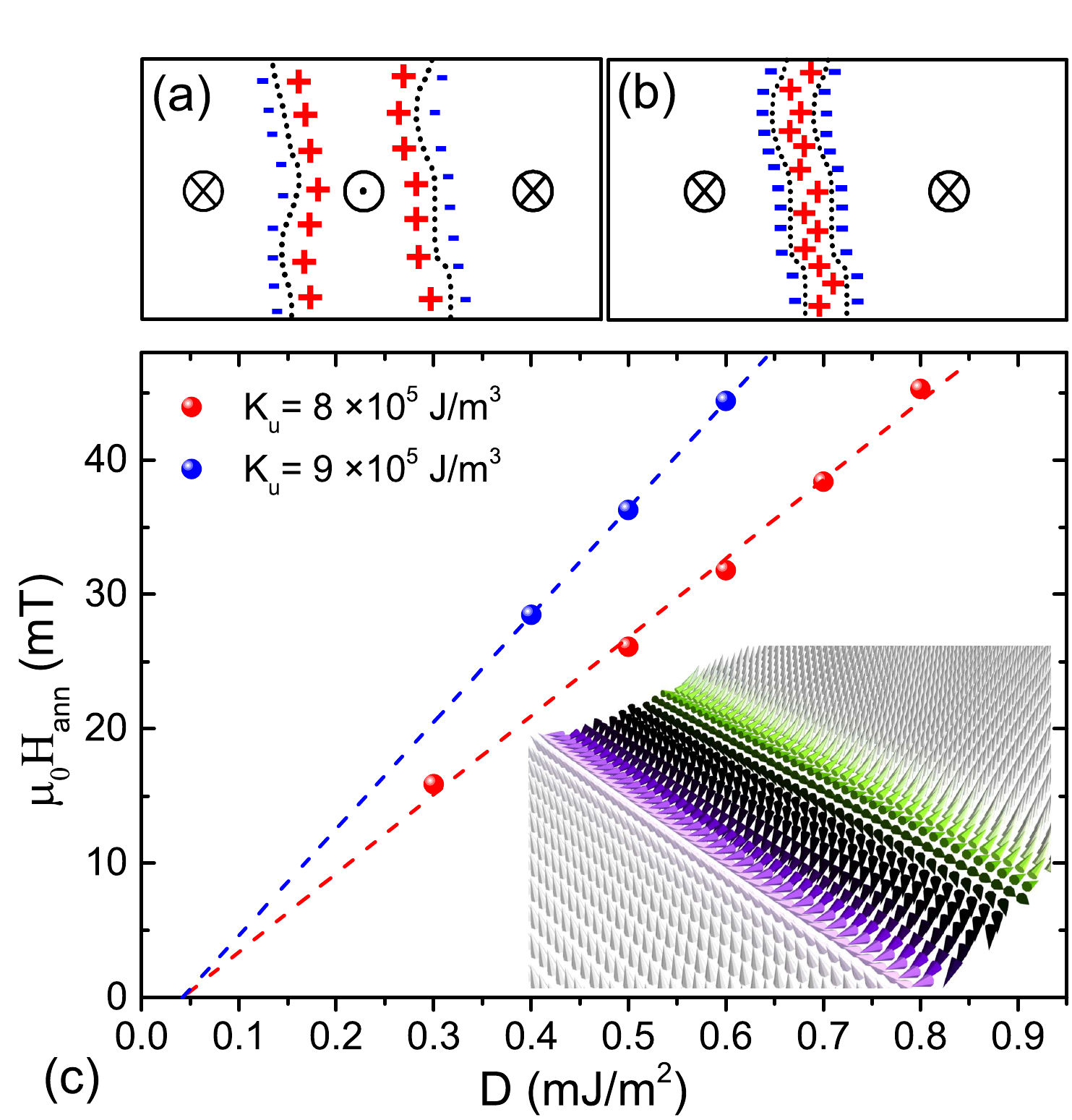}
  \caption{\textbf{Topological prevention of domain wall annihilation by homochirality.} (\textbf{a}) Two 180$^\circ$ N\'{e}el DWs with the same chirality, showing the magnetostatic charges. (\textbf{b}) A composite 360$^\circ$ N\'{e}el DW after application of a small out-of-plane magnetic field that drives them together. The magnetostatic charge prevents the two DWs from annihilation. (\textbf{c}) DW annihilation field as a function of $D$ as calculated by micromagnetic simulations for different uniaxial anisotropies. The inset illustrates the simulated profile of the up-down-up domain including two N\'{e}el walls squeezed together by an external field. \label{DW_Mumax}}
\end{figure}

The nature of the DW chiral handedness was studied in the TEM by varying the objective lens field (which is vertical) to bring the DWs together and observing the subsequent interaction between them. Figure~\ref{DW_squashing}(a)-(c) shows a sequence of images of DWs displaced by a magnetic field. The initial magnetic state shown in figure~\ref{DW_squashing}(a) is imaged before the magnetic field is applied. Figure~\ref{DW_squashing}(b) and (c) show the magnetic state during and after the application of a magnetic field $B_{\mathrm{ext}}=2$~mT. Figure~\ref{DW_squashing}(d) was acquired under the same conditions as figure~\ref{DW_squashing}(c), but for a different in-plane tilt direction as depicted schematically in figure~\ref{DW_squashing}(i), confirming the N\'{e}el state. It can be seen that the up-domain state grows and the down-domain shrinks until the point where the two DWs meet, at which point they merge into 360$^\circ$ DWs. A much larger magnetic field has to be applied in order to annihilate these and fully saturate the film. The formation of 360$^\circ$ DWs takes place when two DWs of the same chirality meet and are unable to annihilate for topological reasons \cite{hubertandschaefer}. Since both DWs have the same chirality, the magnetostatic charge created on either side of the wall prevents the two walls, which are locked and protected by the DMI, from annihilation. This situation is sketched in figure~\ref{DW_Mumax}(a)-(b). It is important to note that \emph{all} of the DW pairs we observed formed 360$^\circ$ DWs when they met, showing that these N\'{e}el walls are homochiral, a situation that is enforced by the presence of an interfacial DMI.

To overcome this topological protection \cite{alejos2015micromagnetic}, a sufficiently strong magnetic field must be applied to overcome the energy barrier and annihilate the $360^\circ$ DW. Measuring the field required to do this has been suggested as a means of experimentally determining the strength of the DMI \cite{Hiramatsu2014}. To collect a statistically significant number of annihilation events, we have mimicked the same experiment with Kerr microscopy measurements in which the up and down domains are easily visible as dark or bright contrast. Domain expansion upon the application of a magnetic field is shown in the Kerr micrographs figure~\ref{DW_squashing}(e)-(g), exhibiting the same behaviour that was observed by the L-TEM. The white lines appearing in figure~\ref{DW_squashing}(g) correspond to the two N\'{e}el walls squeezed together into a $360^\circ$ structure. Since the magnetic field is not large enough to annihilate them, it is possible to pull them apart again by applying a small magnetic field of the opposite polarity as depicted in figure~\ref{DW_squashing}(h). The coercive field $\mu_0H_{\mathrm{c}}=9.0\pm0.2$~mT can be extracted from the Kerr effect hysteresis loop shown inset in figure~\ref{DW_squashing}(j), measured with the magnetic field swept in the range $\pm30$~mT.

In order to measure the annihilation field, the following magnetic field sequence, illustrated by the blue curve in the inset of figure~\ref{DW_squashing}(j), was performed. The film was first saturated at $-30$~mT and then the magnetic field was ramped to a maximum field $\mu_0H_{\mathrm{max}}$ above the coercive field. The magnetic field is then swept back to $-30$~mT and so the switching field $\mu_0H_{\mathrm{sw}}$ can be measured. It is clear from the figure~\ref{DW_squashing}(j) inset that the film now switches at a field that is lower than the coercive field of the major loop, i.e. $\mu_0H_{\mathrm{sw}} \leq \mu_0H_{\mathrm{c}}$.

Figure~\ref{DW_squashing}(j) shows the dependence of the switching field $\mu_0H_{\mathrm{sw}}$ on the maximum field $\mu_0H_{\mathrm{max}}$. The switching field rises linearly until it is as large as $\mu_0H_{\mathrm{c}}$, which occurs at $\mu_0H_{\mathrm{max}}=20.0\pm0.5$~mT. For $\mu_0H_{\mathrm{max}}$ larger than this, $\mu_0H_{\mathrm{sw}} = \mu_0H_{\mathrm{c}}$. The linear rise is caused by a greater proportion of annihilated $360^\circ$ walls, which leads to a lower number of pre-nucleated regions for subsequent reverse domain growth on the reverse portion of the field sweep and a higher switching field. Once all the DWs are annihilated at $\mu_0H_{\mathrm{ann}} \geq 20.0\pm0.5$~mT, the reversal process is governed by reversed domain nucleation and propagation---just as on the major loop---rather than propagation alone. Since the stability of the two N\'{e}el walls in the 360$^\circ$ structure depends on the magnitude of the DMI, the annihilation field is a direct evaluation of the magnitude of the DMI\cite{Hiramatsu2014}.

Micromagnetic simulations were performed to evaluate the effect of the DMI on the DW annihilation field. In this model we used the following material parameters appropriate to Co: saturation magnetization $M_\mathrm{s}=1.1\times10^6$~A/m, exchange stiffness $A=16$~pJ/m, and perpendicular uniaxial anisotropy $K_\mathrm{u}= 9\pm1\times10^5$~J/m$^3$. To investigate this problem we first prepared two DWs in a simulated nanowire of lateral dimensions $256$~nm$\times1024$~nm by introducing an up-down-up domain. The two DWs were set to be 300~nm apart and then brought together by an external field as shown in the inset of figure~\ref{DW_Mumax}(c). Figure~\ref{DW_Mumax}(c) shows the calculated dependence of the annihilation field on $D$ for different anisotropies to include the error of the anisotropy measurement. We find that our measured value of $\mu_0H_{\mathrm{ann}}$ implies that $|D|=0.33\pm0.05$~mJ/m$^2$. Since the micromagnetic simulations were performed at $T=0$~K, and there is a possibility of DW annihilation due to thermal activation, this value is only a lower limit of $D$. The real $D$ could be much be higher, potentially reaching similar value $D\simeq2.2$~mJ/m$^2$ as measured by Pizzini $\textit{et al.}$ \cite{Pizzini_DMI}. In any case, the most pessimistic value of $|D|=0.33\pm0.05$~mJ/m$^2$ is already high enough satisfy the condition $D > 2 N_x \mu_0 M_\mathrm{s}^2 \Delta / \pi$ needed to enforce the N\'{e}el configuration for the DWs in this system \cite{Thiaville_DMI}.

\section*{Summary}

Our L-TEM experiments show that DWs in the Pt/Co/AlO$_x$ system are indeed of the N\'{e}el form, enforced by an interfacial DMI overcoming the preference of the magnetostatic energy term for Bloch walls. The fact that they always form $360^\circ$ structures when pressed together by a field shows that these walls are homochiral, just as is expected in the presence of a DMI. We have implemented the method of annihilating these $360^\circ$ structures with a field in order to estimate the strength of the DMI in this system as being at least $|D|=0.33\pm0.05$~mJ/m$^2$. This confirms the widespread assumption, made to interpret field- or current-driven domain wall motion data, that such walls are of the N\'{e}el type due to an effective field arising from the interfacial DMI. Knowledge of wall structures is not only important in DW dynamics experiments and the design of spintronic technologies based on this phenomenon, such as racetrack memories, where the fact that the DWs are topologically protected from mutual annihilation means that they can be very closely packed, permitting data density. Also, these N\'{e}el type walls, wrapped in a circle, form the boundaries of bubble domains with non-zero skyrmion winding numbers, leading to inertial dynamics \cite{Buttner2015} and with potential for use in skyrmion-based spintronic devices \cite{Nagaosa2013,Fert2013}.

\section*{Methods}

\footnotesize{
Trilayer films of Ta(3.2~nm)/Pt(3~nm)/Co(0.8~nm)/AlO$_x$(3~nm) were deposited by sputtering at base pressure 10$^{-8}$~Torr. The Pt and Co layers were deposited using dc sputtering, whilst the AlO$_x$ layer was deposited by rf sputtering from an oxide target. Si$_4$N$_3$ membranes were used as a substrate, meaning that the samples were immediately ready for transmission electron microscopy studies. The uniaxial anisotropy constant $K_\mathrm{u}$ was measured using a superconducting quantum interference device vibrating sample magnetometer (SQUID-VSM).

The L-TEM images shown here were acquired using an FEI Tecnai T20 TEM operated at 200~kV in Lorentz mode, with the objective lens only weakly excited \cite{benitez2015engineering}. Kerr imaging was carried out in an Evico microscope. The perpendicular magnetic anisotropy of the deposited films was confirmed by polar Kerr microscopy showing square hysteresis loops (e.g. that in figure~\ref{DW_squashing}(i)).

Micromagnetic modelling was carried out using the MuMax3 code \cite{Vansteenkiste2014}. The cell size used was 1~nm $\times$ 1~nm $\times$ 1~nm.
}


\bibliographystyle{naturemag}


\vskip 2cm

\section*{Acknowledgements}
This work was supported by the UK EPSRC (grant numbers EP/I011668/1, EP/I013520/1, EP/K003127/1 and EP/J007110/1), and the Scottish Universities Physics Alliance.

\section*{Author Contributions}
A.H. and A.P.M. deposited the multilayer samples. M.J.B. carried out the L-TEM imaging and image calculation with assistance from S.M. and D.M. A.H. performed the micromagnetic modelling of annihilating domain walls. A.H. performed the Kerr microscopy. A.H., C.H.M. and S.M. wrote the manuscript. All the authors designed the experiments, discussed the data, and reviewed the manuscript.

\section*{Additional Information}
\textbf{Competing financial interests:} The authors declare no competing financial interests.

\textbf{Reprints and permission} information is available online at http://npg.nature.com/reprintsandpermissions.

\end{document}